\documentclass[twoside]{dis09}
\usepackage[latin1]{inputenc}
\usepackage[dvips]{graphicx,epsfig,color}
\usepackage{wrapfig,rotating}
\usepackage{amssymb,amsmath,array}

\newcommand {\pom} {I\!\!P}
\newcommand {\pomsub} {{\scriptscriptstyle \pom}}

\newcommand {\apom} {\alpha_{\pomsub}}

\begin{document}
\title{Electroproduction of Vector Mesons}
\author{Aharon Levy
\thanks{This work was supported in part by the Israel Science Foundation.}
\vspace{.3cm}\\
The Raymond and Beverly Sackler School of Physics and Astronomy\\
Tel Aviv University, 69978 Tel Aviv, Israel \\
}
\maketitle
\begin{abstract}
The energy dependence of the cross section for exclusive
electroproduction of vector mesons is discussed as a way to learn
about the interplay of soft and hard interactions. The question of
determining the scale of these processes is addressed.
\end{abstract}

\section{Introduction}

Exclusive electroproduction of vector mesons (VMs) is a particularly
good process for studying the transition from the soft to the hard
regime of strong interactions~\cite{afs}, the former being well
described within the Regge phenomenology while the latter - by
perturbative QCD (pQCD). The interest in this interplay comes from the
need to understand at which scale a partonic language is
applicable. The exclusive electroproduction of VMs can then be used to
extract information about the generalized parton distributions
(GPDs)~\cite{GPDs}, an essential addition to understanding the
partonic wave-function of the proton.

Among the most striking expectations~\cite{afs} in this transition is
the change of the logarithmic derivative $\delta$ of the cross section
$\sigma$ with respect to the $\gamma^* p$ center-of-mass energy $W$,
from a value of about 0.2 in the soft regime to 0.8 in the hard one,
and the decrease of the exponential slope $b$ of the differential
cross section with respect to the squared-four-momentum transfer $t$,
from a value of about 10 GeV$^{-2}$ to an asymptotic value of about 5
GeV$^{-2}$ when the virtuality $Q^2$ of the photon increases.

\section{Exclusive vector meson photoproduction}

\begin{figure}[h]
\begin{minipage}{0.5\columnwidth}
\hspace{-0.5cm}
\centerline{\includegraphics[width=\columnwidth]{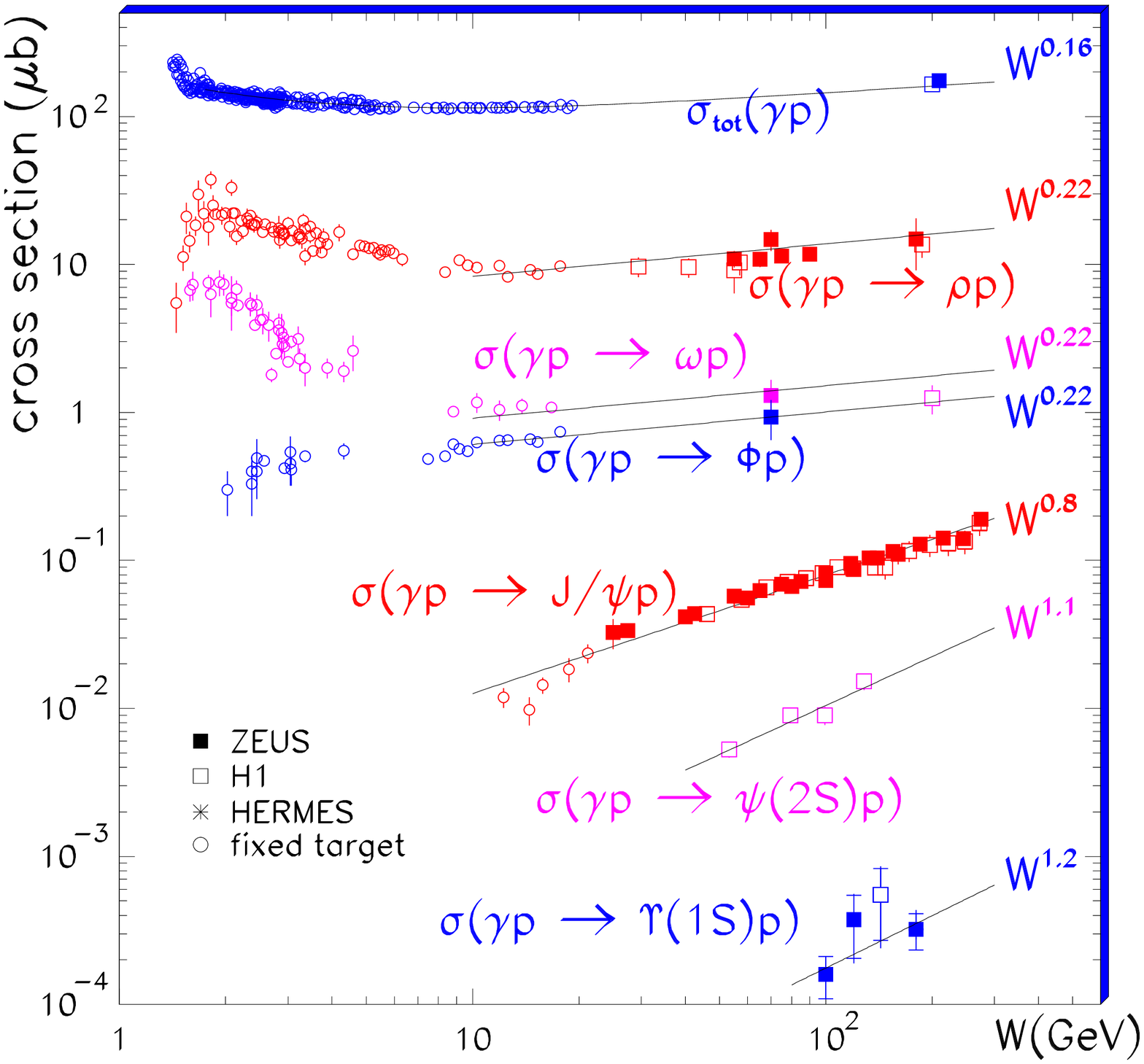}}
\vspace{-0.5cm}
\caption{ Total and exclusive vector meson
photoproduction data, as a function of $W$. The curves are fits of the
form $\sim W^\delta$.  } 
\label{fig:sigvm}
\end{minipage}
\hspace{2mm}
\begin{minipage}{0.5\columnwidth} 
\hspace{-0.5cm}
\centerline{\includegraphics[width=\columnwidth]{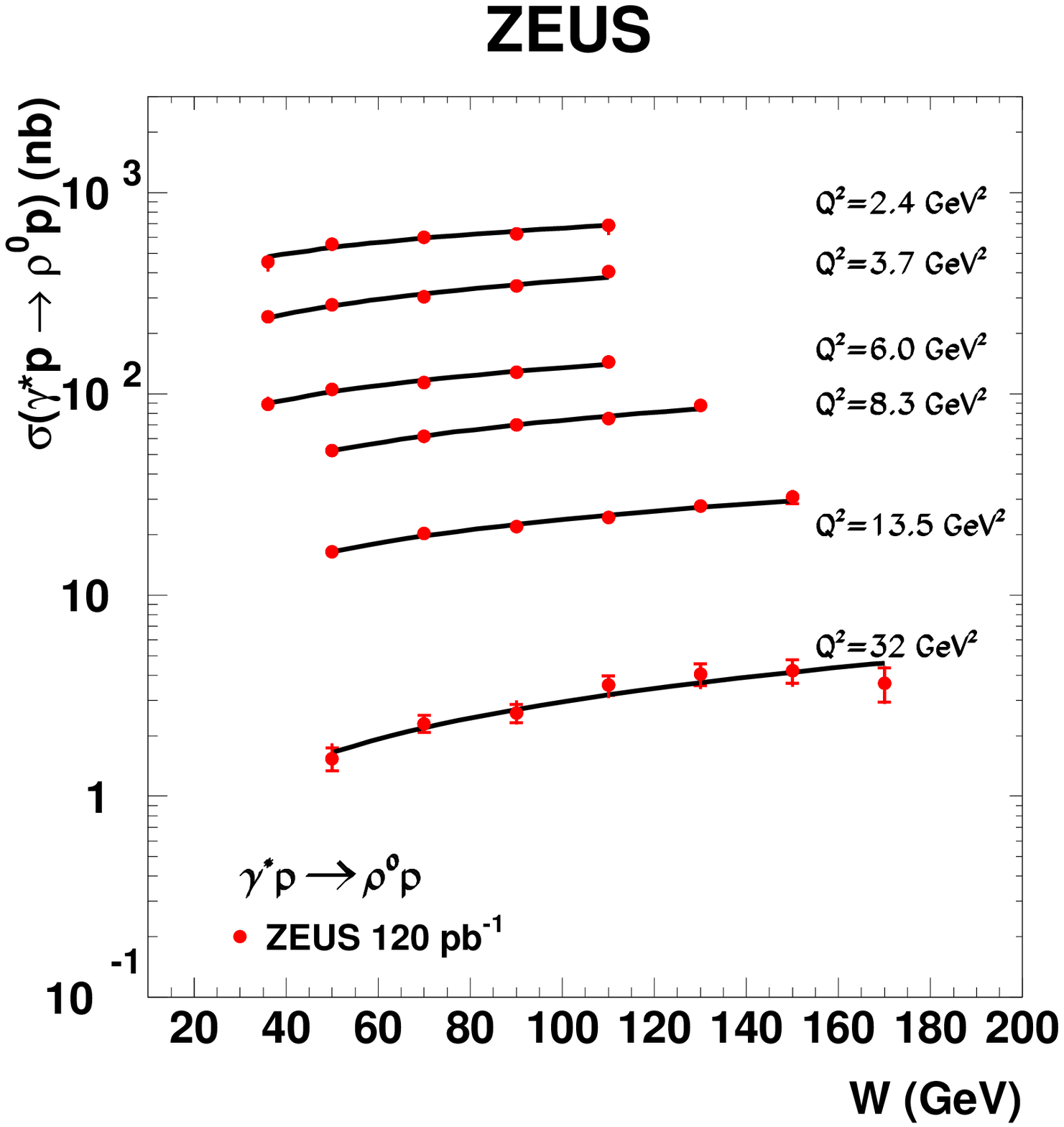}}
\vspace{-0.5cm} 
\caption{\it The $W$ dependence of $\sigma(\gamma^* p\to\rho^0 p)$ for different $Q^2$ values,
as indicated in the figure.  The lines are the result of a fit of the
form $ W^\delta$ to the data.}  
\label{fig:sig-rho} 
\end{minipage} 
\end{figure}

The soft to hard transition can be seen by studying the $W$ dependence
of the cross section for exclusive vector meson photoproduction, from
the lightest one, $\rho^0$, to the heavier ones, up to the
$\Upsilon$. The scale in this case is set by the mass of the vector
meson, as in photoproduction $Q^2$ = 0. Figure~\ref{fig:sigvm} shows
$\sigma(\gamma p \to V p)$ as function of $W$ for light and heavy
vector mesons. 
For comparison, the total photoproduction cross section,
$\sigma_{tot}(\gamma p)$, is also shown. The data at high $W$ can be
parameterised as $W^\delta$, and the value of $\delta$ is displayed in
the figure for each reaction. One sees clearly the transition from a
shallow $W$ dependence for low scales to a steeper one as the scale
increases.


\section{Exclusive vector meson electroproduction}

\begin{figure}[h]
\begin{minipage}{0.5\columnwidth}
\vspace{-5mm} 
\hspace{-0.5cm}
\centerline{\includegraphics[width=1.2\columnwidth]{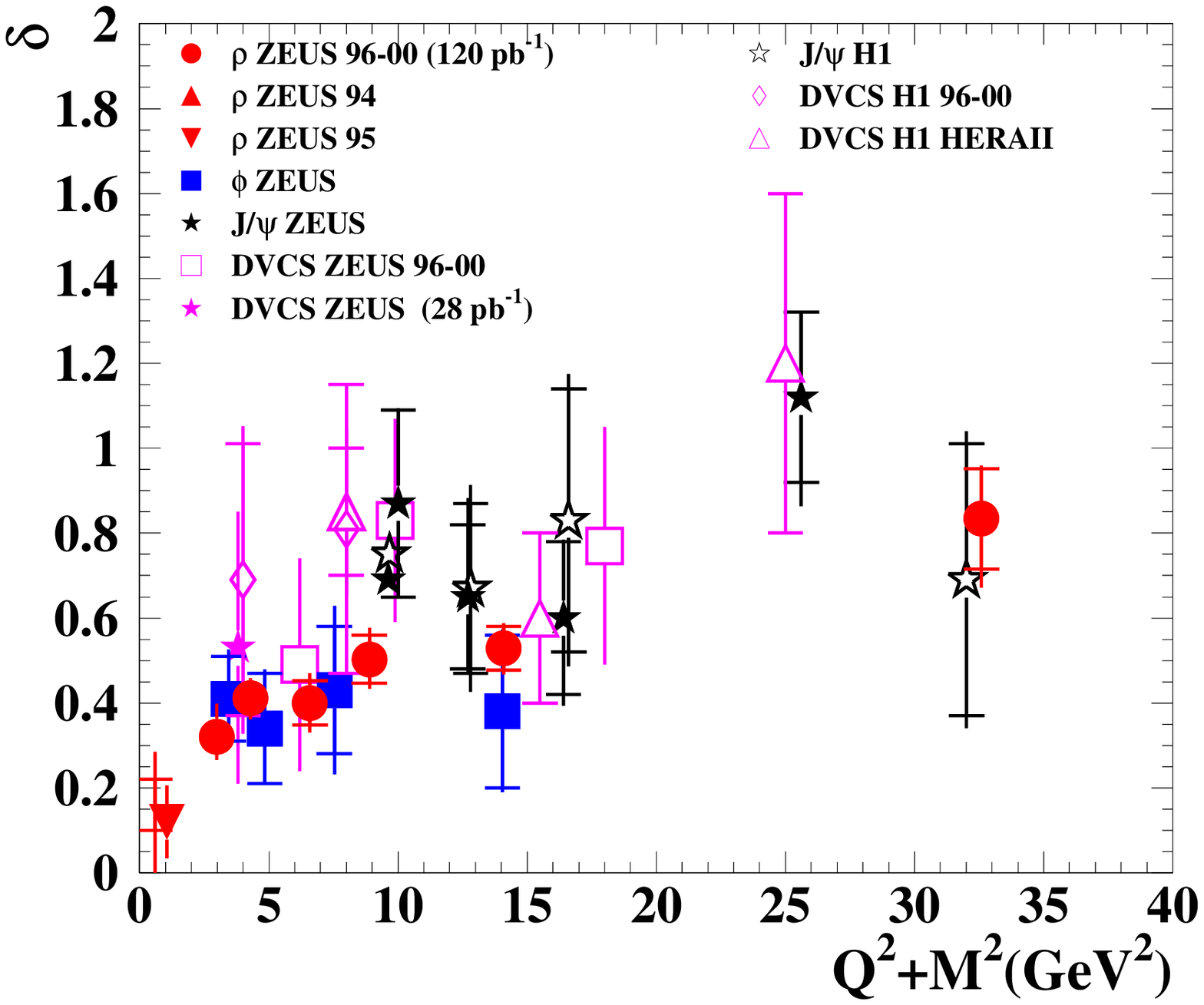}}
\vspace{-0.5cm} 
\caption{\it A compilation of the value of $\delta$
from a fit of the form $W^\delta$ for exclusive vector-meson
electroproduction, as a function of $Q^2+M^2$. It includes also the
DVCS results.}  
\label{fig:del09} 
\end{minipage}
\hspace{2mm} 
\begin{minipage}{0.5\columnwidth} 
\hspace{-0.5cm}
\centerline{\includegraphics[width=\columnwidth]{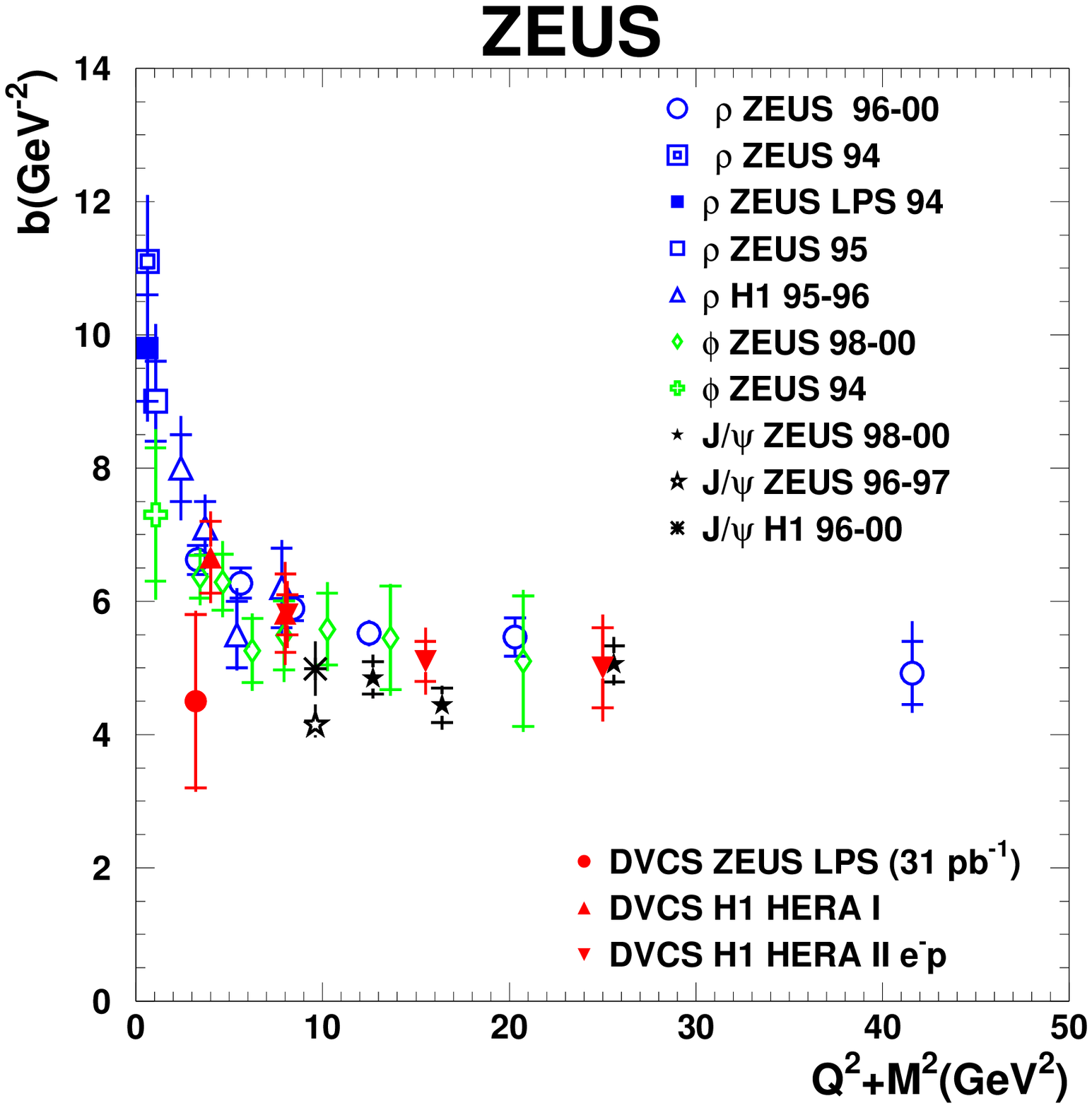}}
\vspace{-0.5cm} 
\caption{ A compilation of the value of the slope $b$
from a fit of the form $d\sigma/d|t| \propto e^{-b|t|}$ for exclusive
vector-meson electroproduction, as a function of $Q^2+M^2$. Also
included is the DVCS result.}  
\label{fig:b09} 
\end{minipage} 
\end{figure}

One can also check this transition by varying $Q^2$ for a given vector
meson.  The cross section $\sigma (\gamma^* p \to \rho^0 p)$ is
presented in Fig.~\ref{fig:sig-rho}~\cite{zeusrho} as a function of
$W$, for different values of $Q^2$. The cross section rises with $W$
in all $Q^2$ bins.  In order to quantify this rise, the logarithmic
derivative $\delta$ of $\sigma$ with respect to $W$ is obtained by
fitting the data to the expression $\sigma \sim W^\delta$ in each of
the $Q^2$ intervals.  The resulting values of $\delta$ from the recent
ZEUS measurement are compiled in Fig~\ref{fig:del09}.  Also included
in this figure are values of $\delta$ from other
measurements~\cite{rho-other} for the $\rho^0$ as well as those for
$\phi$~\cite{zphi,hphi}, $J/\psi$~\cite{zjpsi,hjpsi} and
$\gamma$~\cite{zdvcs,hdvcs} (Deeply Virtual Compton Scattering
(DVCS)). In this case the results are plotted as a function of
$Q^2+M^2$, where $M$ is the mass of the vector meson.  One sees an
approximate universal behaviour, showing an increase of $\delta$ as
the scale becomes larger, in agreement with the expectations mentioned
in the introduction.  The value of $\delta$ at low scale is the one
expected from the soft Pomeron intercept~\cite{dl}, while the one at
large scale is in accordance with twice the logarithmic derivative of
the gluon density with respect to $W$.  
The differential cross section, d$\sigma$/d$t$, has been
parameterised by an exponential function $e^{-b|t|}$ and fitted to the
data of exclusive vector meson electroproduction and also to DVCS. The
resulting values of $b$ as a function of the scale $Q^2+M^2$ are
plotted in Fig.~\ref{fig:b09}. As expected, $b$ decreases to a
universal value of about 5 GeV$^{-2}$ as the scale increases.


\section {The effective scale of vector mesons}

Figures~\ref{fig:del09} and~\ref{fig:b09} might give the
impression that the variable $Q^2+M^2$ can serve as an effective scale
for vector mesons. In the following we will show that this is not the
case and that further study is needed to determine this scale.

One way to study this question is to look at the $W$ dependence of the
cross section ratio $r_V$ of exclusive vector meson electroproduction
to that of the total $\gamma^* p$ one, $r_V \equiv \sigma(\gamma^*p\to
Vp)/\sigma_{tot}(\gamma^*p)$. It was shown~\cite{al-dis02} that using
pQCD arguments this ratio should have, at fixed $Q^2$, a $W$
dependence of the form $r_V \propto W^{2\lambda}/b$, where $\lambda$
is the parameter describing the increase of the proton structure
function $F_2$ with decreasing $x$, $F_2 \propto x^{-\lambda}$. Using
Regge arguments, one obtains that $r_V \propto W^{2(\apom(0)-1)}/b$,
where $\apom(0)$ is the intercept of the Pomeron trajectory. Since
$\lambda = \apom(0)-1$, both approaches predict the same $r_V$
behaviour. The variable $b$ is the exponential slope of the
differential cross section. Both in pQCD and Regge approaches the
ratio $r_V$ rises with $W$. The $W$ dependence is not strongly
affected by $b$ since both for the exclusive electroproduction of
$\rho^0$ and $J/\psi$ shrinkage was found to be
small~\cite{zeusrho,zjpsi}.

\begin{figure}
\centerline{\includegraphics[width=0.55\columnwidth]{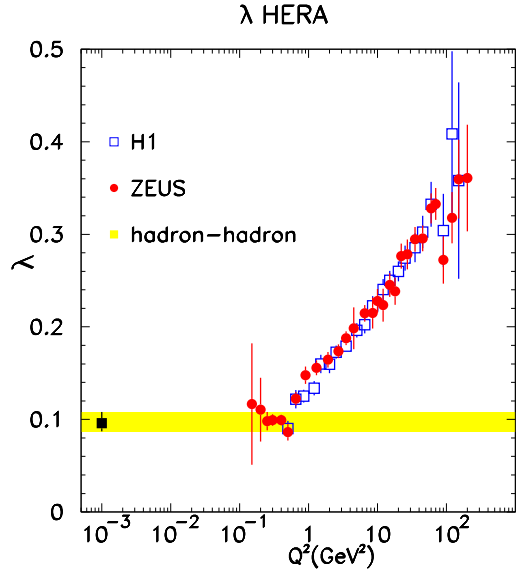}}
\caption{
The values of $\lambda$, obtained by fitting the proton structure
function $F_2$ to the form $\sim x^{-\lambda}$, as a function of
$Q^2$. The shaded band shows $\apom(0)$-1 as obtained from
hadron-hadron total cross section data.
}
\label{fig:lam-hera}
\end{figure}
When calculating the ratio $r_V$ one has to ensure that both cross
sections are taken at the same hard scale, which we denote as
$Q^2_{eff}$. Clearly, for the total inclusive cross section, the
effective scale is $Q^2$. In Fig.~\ref{fig:lam-hera} one can see the
$Q^2$ dependence of $\lambda$ resulting from fitting the $F_2$ data in
the low-$x$ region ($x<0.01$) to the form $F_2 \sim x^{-\lambda}$.
For photoproduction and the low $Q^2$ region, the value of $\lambda$
is in good agreement with that expected from the Pomeron intercept
($\lambda=\apom(0)-1)$. Starting at about $Q^2>$1 GeV$^2$, the value
of $\lambda$ rises logarithmically with $Q^2$. It is of interest to
see if $r_V$ shows the expected $W^{2\lambda}$ behaviour.

It is not clear what is the effective scale for exclusive vector meson
electroproduction. One suggested scale, originally for the
$J/\psi$~\cite{ryskin}, is $Q^2_{eff} = (Q^2 + M^2)/4$. Frankfurt,
Koepf and Strikman~\cite{fks} calculated the effective scale for
$\rho, J/\psi$ and $\Upsilon$, and find that their effective scale for
the $J/\psi$ and for the $\Upsilon$ are significantly larger than that
suggested by~\cite{ryskin}. One can parameterise~\cite{strikman} their
effective scale for the $\rho^0$ as $Q^2_{eff} =
\left(\frac{Q^2}{2.65}\right)^{0.887}$.

As the most precise data at present are those of the exclusive
electroproduction of $\rho^0$~\cite{zeusrho}, we will use these data
to investigate the question of the effective scale.  The ratio
$\sigma(\gamma^* p \to \rho^0 p)/\sigma_{tot}(\gamma^* p)$ as a
function of $W$, for different fixed effective scales, seems to be
constant with $W$ and shows a possible $W$ dependence only at the
highest scale. We have shown earlier that this ratio is expected to
grow with $W$ like $W^{2\lambda}$ in both the pQCD and the Regge
approaches. This would thus indicate that the $\lambda$ values
obtained from this ratio are inconsistent with those obtained in the
inclusive total deep inelastic cross section case.

\begin{figure}[h] 
\begin{minipage}{0.5\columnwidth}
\hspace{-0.5cm} 
\centerline{\includegraphics[width=\columnwidth]{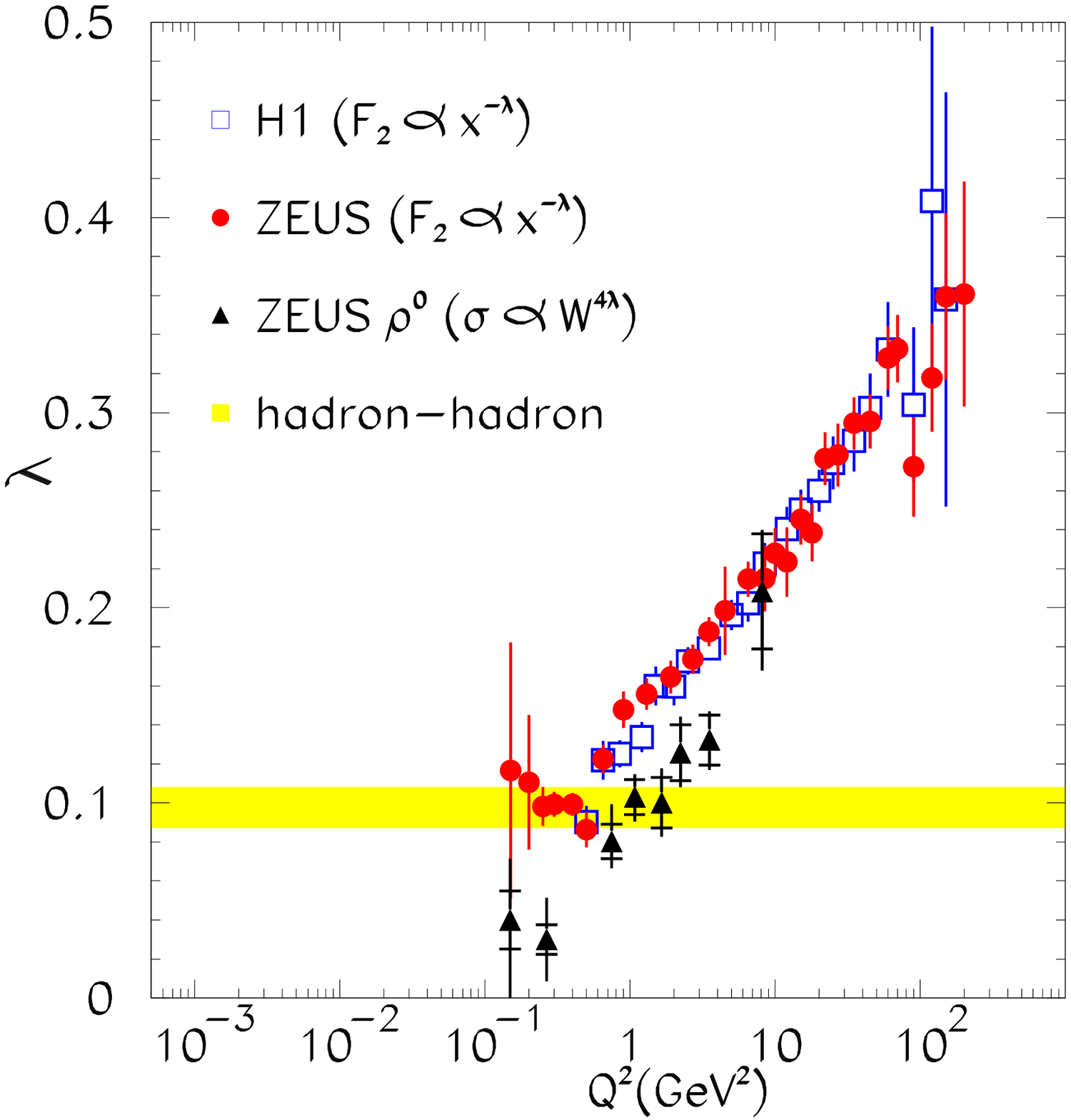}}
\vspace{-0.5cm} 
\caption{ Comparison of the $\lambda$ values obtained
from the exclusive electroproduction of $\rho^0$ with those from the
total inclusive cross section, as a function of $Q^2$ for an effective
scale of the $\rho^0$ of $Q^2_{eff}=(Q^2+M^2)/4$. } 
\label{fig:lam-rho41} 
\end{minipage} 
\hspace{2mm}
\begin{minipage}{0.5\columnwidth} 
\hspace{-0.5cm}
\centerline{\includegraphics[width=\columnwidth]{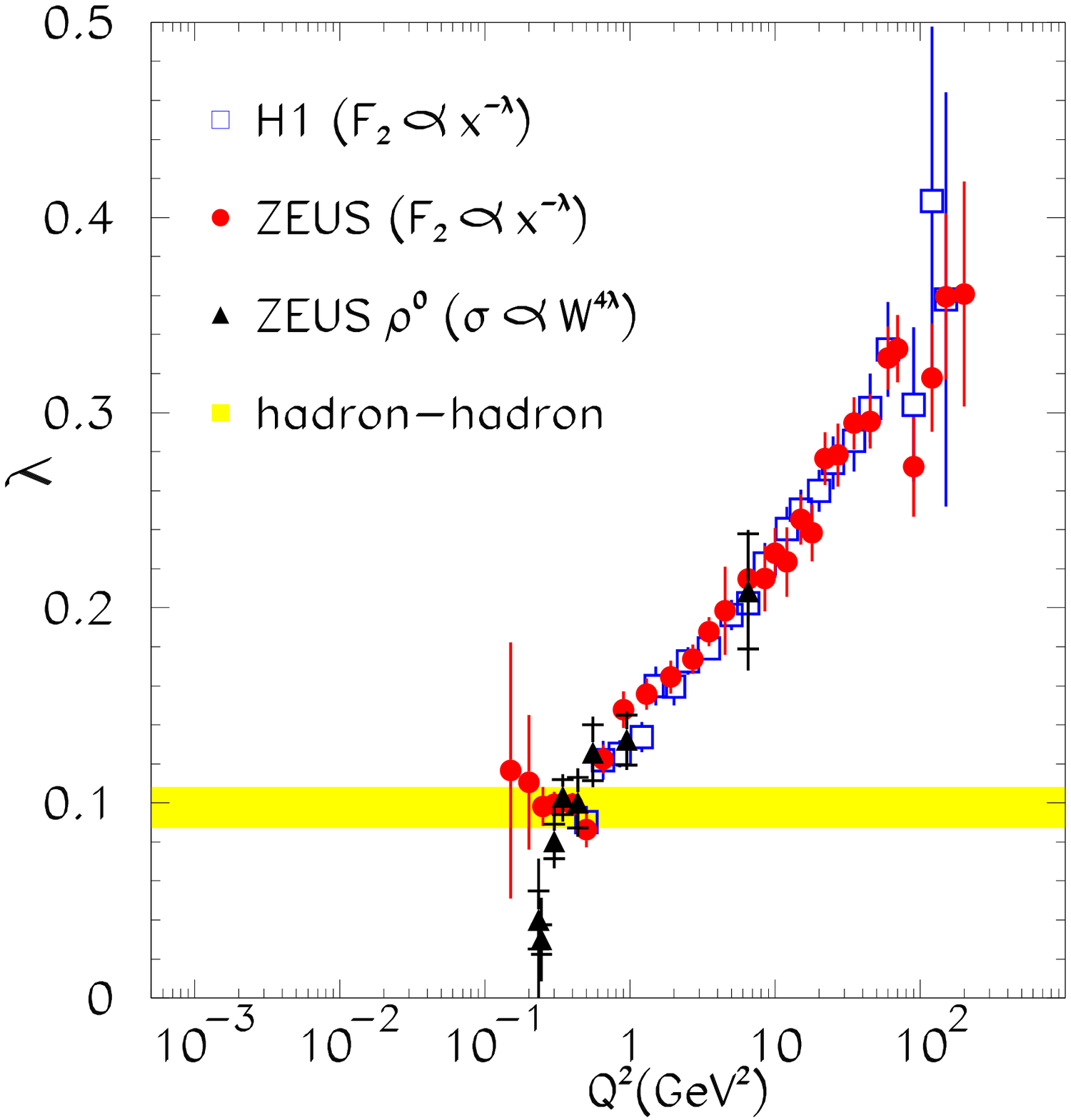}}
\vspace{-0.5cm} 
\caption{The $\lambda$ values obtained from the
exclusive electroproduction of $\rho^0$ together with those from the
total inclusive cross section, as a function of $Q^2$. The effective
scale of $\rho^0$ was chosen so that the two $\lambda$ agree.  } 
\label{fig:lam-rho42} 
\end{minipage} 
\end{figure}

We can in fact compare directly the values of $\delta$ obtained from
the $W$ dependence of the $\rho^0$ cross section, shown in
Fig.~\ref{fig:del09}, to the $\lambda$ values shown in
Fig.~\ref{fig:lam-hera}, keeping in mind that $\delta=4\lambda$. This
is shown in Fig.~\ref{fig:lam-rho41}, for the effective scale
$(Q^2+M^2)/4$.
The $\lambda$ values obtained from the $W$ dependence of the $\rho^0$
are lower than those of the inclusive one for all but the highest
effective scale of the $\rho^0$. This is true also for
$Q^2_{eff}=Q^2+M^2$ as well as for the effective scale suggested
in~\cite{fks}.

One way to find experimentally the right effective scale is to force
the $\lambda$ from the $\rho^0$ to agree with that of the inclusive
data. This is shown in Fig.~\ref{fig:lam-rho42}. The resulting values
of $Q^2_{eff}$ are much smaller than ($Q^2+M_{\rho})$/4 for low $Q^2$
values and can be described by the following ad hoc parameterisation
$Q^2_{eff} = 0.23 e^{0.1Q^2}$.  The fact that the $Q^2_{eff}$ in the
exclusive $\rho^0$ electroproduction is much smaller than $Q^2$ of the
photon might be due to the presence of the convolution of the soft
$\rho^0$ wave-function and the small size longitudinal photon
wave-function. This is a clear sign of the interplay of soft and hard
physics~\cite{afs}.

Unfortunately, the precision of the data on exclusive
electroproduction of $J/\psi$ does not allow a similar study. 

In principle, the question of the effective scale should not be an
issue at all. If we were able to perform calculations in pQCD to all
orders, we would know exactly what is the right scale. However, as
long as we do not yet have a full calculation, precision measurements
of exclusive electroproduction of vector mesons would be helpful to
resolve this problem.

\section{Summary}

The HERA data are a good source to observe the interplay of soft and
hard dynamics, through the study of energy dependences of different
processes.

The process of exclusive electroproduction of vector mesons at high
scales is a good source to study perturbative QCD. It is important to
understand the issue of the effective scale.  This is an essential
step in relating the production of exclusive VMs with the GPDs, the
ultimate source of knowledge about the 3-dimensional partonic
structure of the proton.  For the $\rho^0$, the effective scale is
much smaller than the $Q^2$ of the photon. Better precision
measurements are needed for the $\phi$, $J/\psi$ and DVCS to get a
determination of the effective scale in these processes.

\section{Acknowledgments}
 
It is a pleasure to thank Juan Terron, Claudia Glasman, Cecilia
Uribe-Estrada and Augustin Sabio Vera for organising a most
pleasant and excellent workshop.

\end{document}